\documentclass[showpacs,preprint]{revtex4}
\usepackage{amssymb}
\usepackage{graphicx}
\usepackage{dcolumn}
\usepackage{bm}

\begin{document}

\title{RADIATION EMISSION BY EXTREME RELATIVISTIC ELECTRONS AND PAIR PRODUCTION BY HARD PHOTONS IN A STRONG PLASMA WAKEFIELD}
\author{E.~Nerush and I.~Kostyukov}
\email{kost@appl.sci-nnov.ru}
\affiliation{Institute of Applied Physics, Russian Academy of Science,
  46 Uljanov St. 603950 Nizhny Novgorod, Russia}
\date{\today}

\begin{abstract}

Radiation spectrum of extreme relativistic electrons and a probability of electron-positron pair production by energetic photons in a strong plasma wakefield are derived in the framework of a semiclassical approach. It is shown that that the radiation losses of the relativistic electron in the plasma wakefield scale as $ \propto \varepsilon ^{2 / 3}$ in the quantum limit when the energy of the radiated photon  becomes close to the electron energy, $\varepsilon $. The quantum effects will play a key role in future plasma-based accelerators operating at ultrahigh energy of the electrons. 

\end{abstract}

\maketitle

Crystalline fields along with laser fields are now the most important tools in laboratory strong-field physics. The electric field of atomic strings in a crystal can be up to $10^{11}$~V/cm near a crystallographic axis or plane  \cite{Uggerhoj2005}. When the ultrarelativistic electron with energy about hundred GeV moves along the crystallographic 
axis the electric field can reach the critical electric field $E_{cr} = m^2c^3 / (e\hbar ) \approx 10^{16}\mbox{V / cm}$ in the rest frame of the electron, where $e$ and $m$ is the charge and mass of the electron, respectively, $c$ is the velocity of light, $\hbar $ is Planck's constant. At this field strength quantum effects becomes significant. A number of interesting strong-field phenomena, such as a quantum recoil, pair production, spin flip have been observed in the experiments with crystals (see, e.~g., \cite{Uggerhoj2005} and references therein). Impressive progress in the laser technology during last decades promises to generate even stronger electromagnetic fields than the fields that are achievable in a crystal \cite{Tajima2002}. Nonlinear Compton effect and pair production by nonlinear Breit-Wheeler process have been recently investigated experimentally in collisions of $49$~GeV electron beam with terawatt laser pulses \cite{Bamber1999}. Quantum effects in a strong laser field are now intensively discussed \cite{Mourou2006,Marklund2006}. 

Another example of a strong electromagnetic field, that is  achievable under laboratory conditions, is a strong plasma field. The strong plasma fields can be generated by short intense laser pulse or by short and dense electron bunch propagating in a plasma. The plasma electrons can be completely expelled from interaction region leaving behind a plasma cavity (''bubble``) with uniform ion density \cite{Pukhov2003}. The huge space charge formed due to the electron evacuation generates strong electromagnetic field. The strong plasma wakefield is now considered as a key element of plasma-based accelerators \cite{Leemans2006}, laser-plasma x-ray radiation sources \cite{Kiselev2004}, positron sources based on collisions of plasma-produced hard photons with high-Z materials \cite{Johnson2006}. 

The relativistic electron loaded in the plasma cavity experiences accelerating and focusing forces. The focusing force $F_ \bot \approx m\omega _p^2 r / 2$ is very strong for the ultrarelativistic electron moving in the direction of driver (laser pulse or electron bunch) with $p_z \gg p_ \bot $, where $p_z $ and $p_ \bot $ are  the longitudinal and transverse components of the electron momentum, respectively, $\omega _p = (4\pi e^2 n_0 /m)^{1/2}$ is the electron plasma frequency, $n_0$ is the background plasma density, $r$ is the distance from the electron to the cavity axis. It is assumed that the axially symmetrical cavity and the driver move along $z$-axis. The action of the focusing force leads to the transverse betatron oscillations of the electron about $z$-axis. The betatron frequency is $\omega _b = \omega _p ( 2\gamma )^{ - 1 / 2}$, where $\gamma = \varepsilon / (mc^2)$ is the Lorentz factor related to the energy of the electron, $\varepsilon $. The relativistic electrons undergoing  betatron oscillations emit electromagnetic radiation  \cite{Wang2002}. The spectrum of the radiation becomes synchrotronlike with the critical frequency $\hbar \omega _c = 3\hbar \omega _p^2 r_0 \gamma ^2 / \left( 2c \right) $ when the amplitude of betatron oscillations, $r_0 $, is large $p_ \bot / mc \approx \omega _b r_0 \gamma / c \gg 1$ \cite{Esarey2002}. 

The photon energy increases with increasing the electron energy. When they become close to each other classical description of radiation emission is no longer valid. Quantum effects in strong electromagnetic fields can be characterized by the dimensionless invariants  \cite{Landau} $\chi = e\hbar / (m^3c^5) |F_{\mu \nu } p_\nu | \approx \gamma ( F_ \bot / eE_{cr} )$  and $\Upsilon \approx (\hbar \omega / mc^2 ) (F_ \bot / eE_{cr})$, where $F_{\mu \nu } $ is the field-strength tensor, $p_\mu $ is the particle four-momentum, $\hbar \omega $ is the photon energy. $\chi $ defines the ratio of the electric field strength in the rest frame of thr electron to $E_{cr} $. $\Upsilon $ determines the photon interaction with electromagnetic field. Quantum-electrodynamic effects are important when $\chi \gtrsim 1$ or $\Upsilon \gtrsim 1$. If $\chi \gtrsim 1$ then $\hbar \omega \sim \varepsilon $ and a quantum recoil imposed on the electron by the emitted photon is strong. The invariants can be presented in the form $\chi \approx 10^{ - 6}\gamma $ and $\Upsilon \approx 10^{ - 6} (\hbar \omega / mc^2 )$ for parameters $n_0 \approx 10^{19}$~cm$^{ - 3}$ and $r = 15$~$\mu m$, which are close to the parameters considered in Ref.~\cite{Kiselev2004}. The invariants is close to unity  for the particles with energy $500 $~GeV. For $n_0 = 10^{20}$~cm$^{ - 3}$ the threshold energy, above which quantum effects are decisive, becomes $50 $~GeV.  While the radiation generation in a plasma wakefield has been much studied in the classical limit ($\chi \ll 1$), a little theory for the quantum regime ($\chi \gtrsim 1$) of radiation emission by a relativistic electrons as well as for pair production by decay of an energetic photon in a plasma wakefield currently exists. In this Letter we study these quantum processes.  

The motion of the relativistic electron in the plasma wakefield is semiclassical because the energy level distance of the electron in the plasma wakefield is about $\hbar \omega _b $, that is much less than $\varepsilon $. Yet the radiation emission can be quantum because the energy of the emitted photon can be close to $\varepsilon $. In order to describe the radiation emission in the plasma wakefield it is convenient to use the semiclassical operator method  \cite{Baier1998}. This method strongly simplifies the calculation of the radiation spectrum since the spectrum can be expressed through the parameters of the classical trajectory of the electron. The operator method was applied to calculation of the radiation spectrum in the magnetic field, Coulomb field, crystalline field \cite{Baier1998,Landau}. In the framework of the semiclassical approach the energy radiated by an electron per frequency and per solid angle can be written as follows
\begin{eqnarray}
\frac{d^2W}{d\omega d\Omega _{\bf n} } & = & \frac{e^2 \omega ^2}{4 \pi ^2 c}
\int\limits_{ - \infty }^{ + \infty } {dt_1 } \int\limits_{ - \infty }^{ 
+ \infty } {dt_2 } A \exp (i B),
\label{1}
 \\ 
A & = & \frac{\varepsilon ^2 + \varepsilon '^2}{\varepsilon '^2} \left[{\bm \beta } 
( t_2 ) {\bm \beta}  ( t_1 ) - 1 \right] 
+ \left( \frac{\hbar \omega mc^2}{\varepsilon ' \varepsilon} \right)^2 
 \\ 
B & = & \frac{\varepsilon }{\varepsilon '} \left[ \mathbf{kr} (t_2) - \mathbf{kr} (t_1) 
- \omega (t_2 - t_1 ) \right] ,
\end{eqnarray}
\noindent
where $d\Omega _\mathbf{n} $ is the solid angle around the normal vector $\mathbf{ n} = \mathbf{k} / \left| \mathbf{ k} \right|$, $\omega $ and $\mathbf{ k}$ is the frequency and the wave vector of the emitted photon, respectively, ${\bm \beta } (t) = \mathbf{v}(t) / c$, $\mathbf{v}(t)$ is the electron velocity, $\mathbf{ r} (t)$ is the electron radius-vector, $\varepsilon ' = \varepsilon - \hbar \omega $ is the electron energy after photon emission. The structure of the plasma field is taken into account through the classical trajectory of the relativistic electron defined by $\mathbf{v}(t)$ and $\mathbf{r}(t)$.  

Like in the classical limit  \cite{Kostyukov2003} the main contribution to the integral in Eq.~(3) comes from the neighborhood of the saddle points specified by ${\rm {\bf \dot {v}}} \bot {\rm {\bf k}}$. The radiation of the relativistic electron is confined within a small emission angle about $\theta _e \approx 1 / \gamma $. The length, $l_f $,  of the trajectory part, which gives the main contribution to the integrals in Eq.~(\ref{1}), is of the order of the length, over which the particle is deflected by angle $\theta _e $, $l_f \approx 2\gamma ^2 c \varepsilon ' / (\varepsilon \omega )\approx m c^2 / F_ \bot \approx 2 c^2 / (\omega _p^2 r)$. $l_f $ is also called the formation length (the distance, where an electron creates a photon)  \cite{Uggerhoj2005,Baier1998}. 

The velocity vector of the electron undergoing betatron motion is confined within the deflection angle, $\theta _d \approx \beta _ \bot \approx \omega _b r_0 / c$. The duration of the electron stay inside the plasma cavity, $t_s \approx L_{pc} / \Delta v $, becomes less than the betatron oscillation period, $2 \pi c / \omega_b$ if the electron energy is ultrahigh $\gamma  \gtrsim \gamma _{str} $, where $\Delta v = v - v_d $ is the difference in velocity between the electron and the plasma cavity, $L_{pc} $ is the cavity length. In this case the classical electron trajectory inside the cavity is close to straight line. The plasma cavity velocity is equal to the driver velocity, $v_d$, (velocity of the electron bunch or the group velocity of the laser pulse). Typically $\gamma \gg \gamma _d \gg 1$ then $t_s  \approx 2 \mu /\omega _p  $ and $\gamma _{str} \approx \mu ^2 / \pi ^2 $, where $\gamma _d^{-2} = 1 - v_d^2/c^2$ and $\mu = (\omega _p L_{pc} /c) \gamma _d^2 $.  Thus, we can use the following estimate $\theta _d \approx F_ \bot t_s / (mc\gamma ) \approx  (\omega _p r /c)  \mu / \gamma $ for $\gamma  \gtrsim \gamma _{str} $. In general case of arbitrary $\gamma $ we can write $\theta _d \approx (\omega _b r / c) \min \{1,  \mu (2 / \gamma )^{1/2} \}$ 

The radiation emission is synchrotronlike if $\theta _d \gg \theta _e$. In the opposite limit the radiation emission is dipole. The dipole approximation is valid for the electron trajectory part, where $r \ll r_d $ and $\omega _p r_d /c \approx (2 / \gamma )^{1/2} \max \{1,  ( \gamma / 2)^{1/2} \mu ^{-1} \} $. We limit our calculation to the synchrotron radiation regime since the contribution to the radiation spectrum from the trajectory part, where $r < r_d$, is negligibly small. The reason is that the energy radiated by an electron decreases as $r$  decreases and the transverse radius of the plasma cavity, $R_{pc}$, is much more than $r_d$ because $R_{pc} \sim L_{pc} > c/\omega _p$ \cite{Lu2006}.

The used approach is valid if the focusing field does not significantly change when the electron passes formation length along its trajectory \cite{Baier1998}. Since the focusing force depends only on $r$ \cite{Kostyukov2004} the validity condition related to the field uniformity can be written as follows $l_f \theta _d \ll r  $. This coincides with the condition when dipole radiation is negligible, $r \gg r_d$.  Another validity condition is that the formation length should be much less than the length of the electron trajectory in the plasma wakefield, $l_f \ll c t_s$. The condition fails only for a small part of the electron trajectory, where $r \omega _p /c \ll 1/\mu $. The energy radiated from this trajectory part is small and can be neglected because $\mu \gg 1$ . The electron acceleration in the plasma wakefield can be ignored because the change in the energy of the electron due to the plasma acceleration is much smaller than the electron energy, at which quantum effects are significant, $\Delta \varepsilon \approx F_\parallel c t_s /2 \approx m c^2 \mu ^2 / (2  \gamma _d^2) \ll \varepsilon $ \cite{Kostyukov2006}. 

Integrating Eq.~(\ref{1}) over $t_1 $ and $t_2 $, the angular and spectral distribution of radiation energy can be calculated
\begin{eqnarray}
\frac{d^2W}{d\omega d\Omega _{\rm {\bf n}} } &=& \frac{2e^2\omega ^2\rho 
^2N}{3\pi ^2 c^3 \gamma ^4}\left( {1 + \Psi ^2} \right) 
\nonumber \\ 
&& \times  \left\{ K_{1 / 3}^2 ( x ) \left[ \frac{1}{2}\left( \frac{\hbar \omega }
{\varepsilon '} \right)^2 + \Psi ^2\frac{\varepsilon 
^2 + \varepsilon '^2}{2\varepsilon '^2} \right] \right.
\nonumber \\
&& +  \left. \left( 1 + \Psi ^2 \right) \frac{\varepsilon ^2 + \varepsilon '^2}
{2\varepsilon '^2} K_{2 / 3}^2 \left( x \right) \right\},
\label{2}
\end{eqnarray}
\noindent
where $\Psi = \gamma \psi $, $x = (1 / 3)( \varepsilon / \varepsilon ')( \omega \rho / c)\gamma ^{ - 3}( 1 + \psi ^2)^{3 / 2}$, $\psi $ is the angle between $\mathbf{n}$ and $\mathbf{v} ( t_k )$, $t_k $ is the instance of photon emission corresponding to the saddle point $\mathbf{n} \cdot \mathbf{ \dot v }( t_k ) \approx 0$, $\rho $ is the curvature radius 
of the electron trajectory at $t=t_k $, $N$ is the number of betatron periods that the electron undergoes inside the plasma cavity, $K_\nu (x)$ is the modified Bessel function. 

The classical trajectory of the electron in the plasma wakefield can be approximated as follows  $x = r_0 \cos (\omega _b t)$ \cite{Esarey2002}. Thus, there are $2N$ saddle points which contribute to the integrals in Eq.~(\ref{1}). Like in the classical limit \cite{Kostyukov2003} the curvature radius can be written in the form $\rho \approx (c / \omega _b )(\omega _b^2 r_0^2 / c^2 - \theta ^2\cos ^2\varphi )^{ - 1 / 2}$, where $\psi = \theta \sin \varphi $, $\theta $ is the angle between $\mathbf{n}$ and $z$-axis, $\varphi $ is the azimuthal angle. In the classical limit, $\varepsilon \approx \varepsilon '$ Eq.~(\ref{2}) reduces to the classical formula for the angular and spectral distribution of radiated energy \cite{Kostyukov2003}. If $\gamma > \gamma _{str} $ and the classical trajectory is close to straight line then there is one saddle point. In this case $2N = 1$ and the curvature radius reduces to a form $\rho \approx \gamma m c^2 / F_ \bot  \approx c^2 / (\omega _b^2 r ) $. 

   \begin{figure}
   \includegraphics[width=8cm,clip]{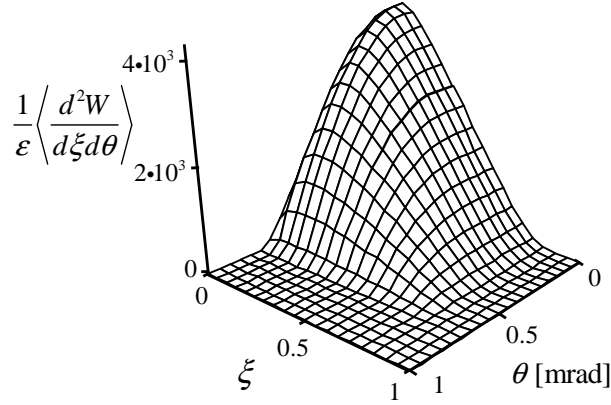}
   \caption
   {
   \label{Fig1}
   The angular and frequency distribution of the normalized energy of radiation per electron, $ \varepsilon ^{-1} \left\langle  d^2 W/ d\xi d\theta \right\rangle $ for the electron beam with energy $300$~GeV during $t_{int} \approx 71.4 \omega _p^{-1}$. The plasma bubble parameters are  $ R_{pc}  \approx 20$~$\mu $m, $n_0=10^{19}$~cm$^{-3}$.
   }
   \end{figure}

It is convenient to introduce the normalized power, $dQ / d\xi = dP / d (\hbar \omega )$, where $Q=P/ \varepsilon $,  $\xi = \hbar \omega / \varepsilon $, $P = dW /dt \approx W/t_s$ is the radiation power, $t_s$ can be expressed in terms of $N$ as follows $t_s \approx  N 2\pi / \omega _b $. Integrating Eq.~(\ref{2}) over solid angle the normalized  spectrum can be calculated
\begin{eqnarray}
 \frac{dQ}{d\xi } & = & \frac{\alpha c}{\sqrt{3} \pi  \lambda _C }
 \frac{\xi}{\gamma } \left[ \left( 1 - \xi + \frac{1}{1 - \xi } \right) K_{2 / 3} ( \delta ) \right.
 \nonumber \\
&& \left. - \int_\delta ^\infty {K_{1 / 3} \left( s \right)} ds \right] ,
\label{3}
\end{eqnarray}
\noindent
where $\alpha = e^2 / \hbar c$ is the fine-structure constant, $\lambda _C = \hbar / m c$ is the Compton wavelength, $\delta = 2\xi / 3(1 - \xi )\chi $. Eq.~(\ref{3}) coincides with the expression for radiation spectrum of an extreme relativistic electron in a crystalline field (see,  e.~g., Eq.~(56) in Ref.~\cite{Uggerhoj2005}). However, in our case  $\chi $ is determined by the plasma wakefield parameters, $\chi \approx \hbar \gamma \omega _p^2 r / (2 mc^3 )$

The expression for the total radiated power can be calculated by integrating of Eq.~(\ref{3}) over the photon energy. Like in the theory of the electron radiation in magnetic field or in crystalline field \cite{Uggerhoj2005,Landau} the radiated power is $P / P_{cl} \approx 1 - 55 \sqrt 3 \chi / 16 + 48\chi ^2$ in the classical limit with account of quantum corrections, while in the limit $\chi \gg 1$ it can be expressed as follows $P / P_{cl} \approx 1.2\chi ^{ - 4 / 3}$, where $P_{cl} \approx e^2\gamma ^2\omega _p^4 r_0^2 / 12c^3$ is the classical radiated power of an electron in plasma wakefield \cite{Esarey2002}. It follows from the obtained expressions that the electron radiation losses in the plasma wakefield scale as $ \propto \varepsilon ^{2 / 3}$ in the quantum limit ($\chi \gg 1$) whereas the radiation losses scale as $ \propto \varepsilon ^2$ in the classical limit. 
     
The angular and frequency distribution of the normalized energy radiated per electron, $ \varepsilon ^{-1} \left\langle  d^2 W/ d\xi d\theta \right\rangle $, during interaction time $t_{int} = 72 \omega _p^{-1}$ is shown in Fig.~\ref{Fig1} for the electron beam with energy $300$~GeV. The distribution can be calculated by averaging of Eq.~(\ref{2}) over $\varphi$ and $r_0$. The structure of the plasma wakefield is similar to that considered in Ref.~\cite{Kiselev2004} where radiation emission was modeled in the classical limit. The plasma bubble parameters are  $R_{pc}  \approx 20$~$\mu $m, $n_0=10^{19}$~cm$^{-3}$. $\chi \approx 0.8$ for the electron with $r \approx R$ and the classical theory of radiation emission is no longer valid. The energy damping length of this electron because of radiation emission, $l_r \approx c \varepsilon / P $, is about five times more than $l_r$ estimated by the classical theory. It is seen from Fig.~\ref{Fig1} that the spectrum peaks near $40$~GeV whereas the classical theory predicts the spectrum maximum near $80$~GeV.  Like in the classical limit \cite{Kiselev2004} the radiation angle is close to the electron deflection angle $\theta _d \approx 0.7$~mrad that is in good agreement with Fig.~\ref{Fig1}.

   \begin{figure}
   \includegraphics[width=7cm,clip]{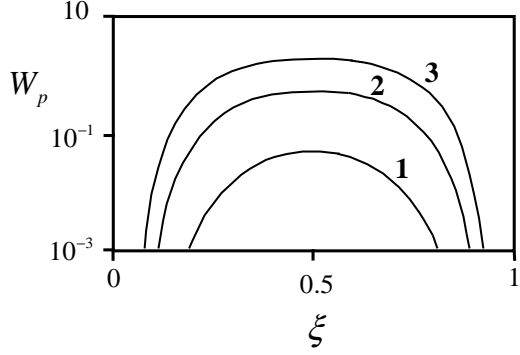}
   \caption
    {
        \label{Fig2}
The energy distribution of the positrons (electrons) created by the photon with energy $300$~GeV in the plasma bubble for different values of $r$ (1) $r=10$~$\mu$m, (2) $r=15$~$\mu$m, (3) $r=20$~$\mu$m, respectively. The parameters of the plasma bubble are $L_{pc} \approx 2R_{pc}  \approx 40$~$\mu $m, $n_0=10^{19}$~cm$^{-3}$ and $\gamma _d \approx 10$. 
    }
   \end{figure}

An electron-positron pair can be created by an energetic photon in strong electromagnetic fields \cite{Landau}. Therefore, photon emission by relativistic electrons can cause inverse process of pair production by a hard photon in a strong plasma field. The pair production is related to photon emission by crossing channel. The probability of pair production can be calculated by the semiclassical method used above \cite{Baier1998}. The probability per unit time is 
\begin{equation}
\label{pair}
W_p = \frac{C}{6\sqrt 3 \pi }\int\limits_1^\infty {\frac{8z + 1}{z^{3 / 2}\sqrt {z - 1} }} K_{2 / 3} \left( {\frac{8z}{3\Upsilon }} \right)dz,
\end{equation}
\noindent
where $C = \alpha m^2c^4 / (\hbar ^2\omega )$. It follows from Eq.~(\ref{pair}) that $W_p $ is determined by invariant $\Upsilon \approx (1/2) (\hbar \omega / mc^2) (\hbar \omega _p / mc^2 ) (r \omega _p /c) $. In the limit $\Upsilon \ll 1$ the probability is exponentially small $W_p \approx 0.2C\Upsilon \exp ( - 8 / 3\Upsilon )$, while in the limit $\Upsilon \gg 1$ it is $W_p \approx 0.4C\Upsilon ^{2 / 3}$. The pair formation length can be defined as a length, it takes to deflect a created positron (electron) to an angle $\theta _e$ with respect to photon wave vector. It is equal to $l_f$ that reflects the crossing symmetry of pair production and photon emission. The validity conditions of Eq.~(\ref{pair}) are identical to that discussed above for Eq.~(\ref{1}).

The energy distribution of positrons (electrons) created by a photon with energy $300$~GeV in the plasma bubble is shown in Fig.~\ref{Fig2} for different values of $r$, where $r$ is the distance between the photon and $z$-axis. The photon propagates in the direction of $z$-axis. The plasma bubble parameters are  $L_{pc} \approx 2R_{pc}  \approx 40$~$\mu $m, $n_0=10^{19}$~cm$^{-3}$ and $\gamma _d \approx 10$.  It is seen from Fig.~\ref{Fig2} that the distribution function peaks at $\varepsilon = \hbar \omega /2$. The positron energy spread  increase as $r$ increase. The probability of the pair production by the photon crossed the bubble with $r=20$~$\mu$m is about $1.5$. It is of the order of the pair production probability by the photon passed the same distance ($\approx 0.8$~cm) in the field maximum of Ge crystal cooled to $100$~K \cite{Baier1998}. 

In Conclusion we have calculated the radiation spectrum of energetic electrons and the probability of pair production by a hard photon in a strong plasma wakefield. Like crystals \cite{Uggerhoj2005} a plasma can be a very efficient radiator for high energy electrons as well as an intense positron source. Moreover, a plasma wakefield has some advantages over crystalline and laser fields because it is a large-scale structure by comparison. The intense crystalline field is located in very narrow layer ($\lesssim 10^{-4}$~$\mu $m) about the crystallographic axis and the intense laser radiation varies over the laser wavelength ($\lesssim 1$~$\mu $m). The width of the high energy electron beam is typically ten microns and more \cite{Johnson2006} that is close to the plasma cavity size. This provides more efficient interaction of the electron beam with strong plasma wakefield than with strong crystalline and laser fields. In addition, the ponderomotive force of the intense laser pulse pushes out the electrons from the region with strong laser field, whereas the plasma wakefield focuses the electron beam. It should be noted that the plasma wakefield can be used as a high-gradient accelerating structure \cite{Leemans2006,Johnson2006} that open possibility for developing of the compact photon and positron source. 

Radiation losses in the plasma-based accelerators increase as $\varepsilon ^2 $ in the classical limit and strongly affect dynamics of energetic electrons \cite{Kostyukov2006,Michel2006}. We found that the radiation losses scale as $ \propto \varepsilon ^{2 / 3}$ in the quantum limit, which will be achievable in future plasma-based high-energy accelerators. Other quantum effects (effects accompanying collisions of charged particles \cite{Berezhiani1992}, vacuum breakdown \cite{Bulanov2004,Mourou2006}, photon splitting, etc.) can be important in a very strong plasma field.

\begin{acknowledgments}
This work has been supported by Russian Foundation for Basic Research (Grant No 04-02-16684, No 05-02-17367). 
\end{acknowledgments}

\end{document}